\documentclass{jpp}

\usepackage{graphicx}
\usepackage{natbib}
\usepackage{amsmath}
\usepackage{color}

\ifCUPmtlplainloaded \else
  \checkfont{eurm10}
  \iffontfound
    \IfFileExists{upmath.sty}
      {\typeout{^^JFound AMS Euler Roman fonts on the system,
                   using the 'upmath' package.^^J}%
       \usepackage{upmath}}
      {\typeout{^^JFound AMS Euler Roman fonts on the system, but you
                   dont seem to have the}%
       \typeout{'upmath' package installed. JPP.cls can take advantage
                 of these fonts, if you use 'upmath' package.^^J}%
      }
  \else
  \fi
\fi

\ifCUPmtlplainloaded \else
  \checkfont{msam10}
  \iffontfound
    \IfFileExists{amssymb.sty}
      {\typeout{^^JFound AMS Symbol fonts on the system, using the
                'amssymb' package.^^J}%
       \usepackage{amssymb}%
       \let\le=\leqslant  
       \let\ge=\geqslant  
      }{}
  \fi
\fi 

\ifCUPmtlplainloaded \else
\IfFileExists{amsbsy.sty}
             {\typeout{^^JFound the 'amsbsy' package on the system, using it.^^J}%
     \usepackage{amsbsy}}
    {}
\fi

\newsavebox{\astrutbox}
\sbox{\astrutbox}{\rule[-5pt]{0pt}{20pt}}

\newcommand{\Alfven}{Alfv\'{e}n }
\newcommand{\Alfvenic}{Alfv\'{e}nic }
\newcommand{\T}[1]{{\tt #1}} 
\newcommand{\V}[1]{\mathbf{#1}} 
\newcommand{\xhat}{\mbox{$\hat{\mathbf{x}}$}} 
\newcommand{\yhat}{\mbox{$\hat{\mathbf{y}}$}} 
\newcommand{\zhat}{\mbox{$\hat{\mathbf{z}}$}} 
\newcommand{\bhat}{\mbox{$\hat{\mathbf{b}}$}} 

\title[Magnetic Field Line Wander]{The Development of Magnetic Field
  Line Wander by Plasma Turbulence}

\author[ G.~G.~Howes and  S.~Bourouaine]%
       {G\ls R\ls E\ls G\ls O\ls R\ls Y\ns G.\ns H\ls O\ls W\ls E\ls S \ns and\ns
         S\ls O\ls F\ls I\ls A\ls N\ls E\ns B\ls O\ls U\ls R\ls O\ls U\ls A\ls I\ls N\ls E}

\affiliation{Department of Physics and Astronomy,
University of Iowa, Iowa City IA 54224, USA}

\pubyear{2010}
\volume{650}
\pagerange{119--126}
\date{?; revised ?; accepted ?. - To be entered by editorial office}
\begin{document}

\maketitle

\begin{abstract}
Plasma turbulence occurs ubiquitously in space and astrophysical
plasmas, mediating the nonlinear transfer of energy from large-scale
electromagnetic fields and plasma flows to small scales at which the
energy may be ultimately converted to plasma heat. But plasma
turbulence also generically leads to a tangling of the magnetic field
that threads through the plasma. The resulting wander of the magnetic
field lines may significantly impact a number of important physical
processes, including the propagation of cosmic rays and energetic
particles, confinement in magnetic fusion devices, and the fundamental
processes of turbulence, magnetic reconnection, and particle
acceleration.  The various potential impacts of magnetic field line
wander are reviewed in detail, and a number of important theoretical
considerations are identified that may influence the development and
saturation of magnetic field line wander in astrophysical plasma
turbulence. The results of nonlinear gyrokinetic simulations of
kinetic Alfv\'en wave turbulence of sub-ion length scales are
evaluated to understand the development and saturation of the
turbulent magnetic energy spectrum and of the magnetic field line
wander.  It is found that turbulent space and astrophysical plasmas
are generally expected to contain a stochastic magnetic field due to
the tangling of the field by strong plasma turbulence. Future work
will explore how the saturated magnetic field line wander varies as a
function of the amplitude of the plasma turbulence and the ratio of
the thermal to magnetic pressure, known as the plasma beta.
\end{abstract}

\begin{PACS}
\end{PACS}

\section{Introduction}
Turbulence remains one of the great unsolved problems of classical
physics.  Throughout the universe, from distant galaxy clusters to our
own heliosphere, 99\% of baryonic matter occurs in the plasma state, and
these plasmas are nearly always found to be magnetized and
turbulent. On the frontier of plasma physics research is the effort to
understand how turbulence affects the evolution of any system in which
it arises, from terrestrial settings to distant regions of the
universe.  Plasma turbulence mediates the conversion of the energy of
plasma flows and magnetic fields at large scales to plasma heat, or
other forms of particle energization. Turbulence may also be a key
ingredient in the acceleration of high energy particles at
collisionless shocks, and magnetic irregularities caused by turbulence
affect the propagation of energetic particles, such as cosmic rays in
the Galaxy and solar energetic particles in the heliosphere.  The
physics of magnetic reconnection may be fundamentally altered in a
turbulent medium.  Turbulence enhances the loss of angular momentum
from accretion disk plasmas, enabling the fueling of black holes and
other compact objects.  In terrestrial laboratories, turbulence limits
the efficiency of proposed magnetically confined fusion devices by
enhancing the transport of heat and particles across the confining
magnetic field.

\begin{figure}
  \centerline{
\resizebox{\textwidth}{!}{\includegraphics{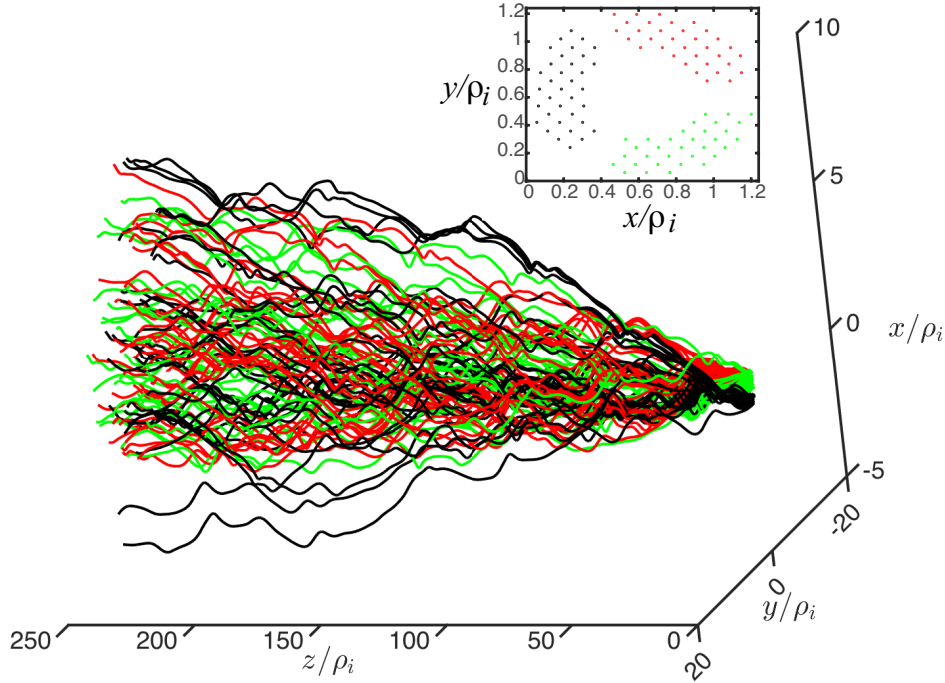}}
  }
\caption{Three-dimensional plot of the spreading and tangling of
magnetic field lines in a driven, nonlinear gyrokinetic simulation of
plasma turbulence relevant to the solar wind.  Field lines within four
small regions at $z=0$ are colored red, green,
blue, and black. From an initially straight magnetic field,
continually driven turbulence, possibly in combination with magnetic
reconnection, leads to a stochastically tangled magnetic field. Note
the scale of the $z$-axis is compressed in this plot, so the domain is, in
fact, highly elongated.
\label{fig:tangle}}
\end{figure}

Although studies of astrophysical turbulence generally focus on how
turbulence mediates the conversion of energy from one form to another,
plasma turbulence also naturally generates a tangled magnetic
field. For example, consider a quiescent, incompressible
magnetohydrodynamic (MHD) plasma embedded with a straight and uniform
magnetic field.  If one drives finite-amplitude \Alfven waves in both
directions along that magnetic field, nonlinear interactions among
those counterpropagating \Alfven waves will lead to a turbulent
cascade of fluctuation energy to small scales
\citep{Kraichnan:1965,Similon:1989,Sridhar:1994,Goldreich:1995,Maron:2001,Howes:2012b,Howes:2013a,Nielson:2013a,Howes:2013b,Drake:2013}. In
the process, the magnetic field becomes increasingly tangled, taking
on a stochastic appearance, as shown by the rendering of magnetic
field lines from a driven, gyrokinetic simulation of plasma turbulence
in figure~\ref{fig:tangle}. Here, magnetic field lines passing through
four small regions at $z=0$ are given distinct colors. As the field
lines are followed through the simulation domain, they spread out and
become increasingly tangled up with field lines initially from other
regions of the plasma.  This phenomenon of \emph{magnetic field line
  wander} arises due to the nonlinear interactions that mediate the
turbulent cascade of energy to small scales, likely in combination
with the process of magnetic reconnection.

The inevitable wandering of the magnetic field in turbulent plasmas
affects a number of other physical processes, with important
unanswered questions regarding its impact on energetic particle
propagation in astrophysical and fusion plasmas, the cascade of energy
in plasma turbulence, particle acceleration, and magnetic
reconnection. Most of the existing studies that attempt to assess the
effect of magnetic field line wander on other physical processes, as
reviewed below in Section~\ref{sec:impact}, have used a turbulent
magnetic field derived from a simplified model or other analytical
prescription rather than a turbulent magnetic field generated by
direct numerical simulations. Here we advocate a more fundamental
approach, using direct numerical simulations to develop a quantitative
understanding of how a magnetic field becomes tangled in plasma
turbulence as a function of the parameters of the plasma and the
turbulence.  Theoretical considerations of the tangling of a magnetic
field in plasma turbulence are presented in Section~\ref{sec:theory},
including the fundamental parameters upon which the development of
magnetic field line wander is likely to depend. Finally, we present
some initial quantitative results on the development of magnetic
stochasticity and the separation of field lines using nonlinear
gyrokinetic simulations that accurately resolve the kinetic
microphysics of collisionless magnetic reconnection.

\section{Impact of Magnetic Field Line Wander}
\label{sec:impact}
Magnetic field line wander arising from plasma turbulence impacts
important plasma physics processes that govern the evolution of a wide
range of important space, astrophysical, and laboratory plasma
systems. We review below previous investigations of how the wandering
of the magnetic field affects the propagation of energetic particles
in astrophysical and fusion plasmas, the cascade of energy in plasma
turbulence, the acceleration of particles at collisionless shocks, and
the physics of magnetic reconnection.

\subsection{Propagation of Cosmic Rays and Energetic Particles}

The attempt to understand and predict the propagation of cosmic rays
through the interplanetary and interstellar magnetic fields has been a
major driver of research into the effect of turbulence on the tangling
of the magnetic field. In a seminal early paper,
Jokipii \citep{Jokipii:1966} performed the first detailed quasilinear
statistical calculation of the motion of charged particles in a
spatially random magnetic field, establishing a quantitative
connection to the turbulent power spectrum of magnetic field
fluctuations. Subsequent work conceptually explained the observed
spreading of solar energetic particles from an active region over
180$^\circ$ in solar longitude as a consequence of a
magnetic-field-line random walk due to turbulent magnetic field
fluctuations \citep{Jokipii:1968}.

The scattering and acceleration of cosmic rays by a spectrum
of \Alfven waves with strictly parallel wavenumbers was treated
analytically by Schlickeiser \citep{Schlickeiser:1989}, and was later
extended to include the interaction with fast-mode waves in a low beta
plasma \citep{Schlickeiser:1998}. A nonlinear diffusion theory of the
stochastic wandering of magnetic field lines, developed by
Matthaeus \emph{et al.} \citep{Matthaeus:1995}, lead to the expectation
of diffusive field line wandering in the perpendicular direction.

By the mid-1990s, numerical modeling of the wandering of magnetic field
lines began to be widely used, including test particle calculations of
energetic particle transport along those turbulent magnetic fields.
These efforts require, as input, a model of the spectrum of magnetic
fluctuations generated by the plasma turbulence, and a wide variety of
such turbulent magnetic field models have been used.  Sophisticated
numerical field-line following algorithms and complementary analytical
approaches have been used to study realizations of slab turbulence
models with
$\delta \V{B}(z)$ \citep{Schlickeiser:1989,Shalchi:2007a,Shalchi:2007b,Shalchi:2010a},
2D turbulence models with
$\delta \V{B}(x,y)$ \citep{Shalchi:2007a,Shalchi:2007b,Guest:2012},
composite models including slab plus 2D components with
$\delta \V{B}= \delta \V{B}(z)
+ \delta \V{B}(x,y)$ \citep{Bieber:1996,Giacalone:1999,Shalchi:2007a,Shalchi:2007b,Qin:2013},
and full 3D models with $\delta \V{B}(x,y,z)$, including both
isotropic
\citep{Zimbardo:1995,Giacalone:1999,Ragot:2011a,Shalchi:2010b}
and anisotropic distributions of magnetic
fluctuations \citep{Chandran:2000,Zimbardo:2000,Zimbardo:2006,Shalchi:2013,Ruffolo:2013}.

These various investigations, conducted by a wide range of
researchers, have often found conflicting results. The mean square
displacement $\langle (\delta r)^2\rangle$ between two magnetic field
lines, as they are followed along the  magnetic field
line a distance  $l$, is often modeled by the power-law form
\begin{equation}
  \langle (\delta r)^2\rangle \propto |l|^p.
\end{equation}
For different turbulent magnetic field models, or even  a
variation of the parameters within a single model, the resulting
magnetic field line wandering is sometimes found to be sub-diffusive
($p <1$) and other times found to be super-diffusive ($p>1$), and yet
other studies recover a standard diffusive behavior ($p=1$).  Based on
a broad reading of the literature, the results of the analytical and
numerical modeling appear to be rather sensitive to the parameters and
properties of the turbulence model chosen. For example, analytical
modeling of anisotropic 3D turbulence (with $k_\perp \gg k_\parallel$)
suggests the field line wandering is diffusive
($p=1$) \citep{Shalchi:2013}, in contrast to the anomalous diffusion
($p \ne 1$) often found using other turbulent models. At present, the
wandering of magnetic field lines, and its impact on the propagation
of energetic particles, remains an active area of research.

Improved modeling of the propagation of energetic particles in
turbulent magnetic fields can have a significant impact on our
technological infrastructure, with serious implications for
national security.  Our society is increasingly dependent on
space-borne assets, such as GPS navigation and communication
satellites, so the prediction of severe space weather events in
near-Earth space has become critically important for the protection of
these assets.  Solar energetic particle (SEP) events, in which a
violent event on the surface of the sun spews high-energy electrons
and protons into the heliosphere, represent a threat to both robotic
and human assets in space. This prompts an urgent need to develop the
ability to predict whether high-energy particles from a particular SEP
event will reach the position of a potentially susceptible
spacecraft. A predictive capability requires understanding how the SEP
particles propagate through the turbulent interplanetary magnetic
field. For example, on 3 NOV 2011, an SEP event erupted on the far
side of the solar surface, spewing out energetic protons and electrons
that were measured at the STEREO A, SOHO, and STEREO B spacecraft,
covering more than 200$^\circ$ of solar longitude
\citep{Richardson:2014}. This wide longitudinal spread of significant
SEP particle fluxes is not satisfactorily predicted by existing
models of SEP propagation.  An improved understanding of the magnetic
field line wander in the turbulent interplanetary magnetic field, as a
function of the plasma and turbulence parameters, is necessary to
develop a reliable predictive capability.

In addition to studies of energetic particle propagation, thermal
conduction in astrophysical plasmas, such as that occurring in
galaxy-cluster cooling flows, has also been found to be strongly
affected by a stochastic magnetic field \citep{Chandran:1998}.

\subsection{Magnetic Confinement Fusion}

Coincident with the earliest studies on the propagation of energetic
particles in space and astrophysical plasmas were complementary
studies of anomalous electron heat transport in tokamaks of the
magnetic confinement fusion program. In tokamak plasmas,
gradient-driven instabilities generate turbulent fluctuations in the
confining magnetic field.  It has been proposed that the distortion of
magnetic flux tubes as they are mapped along the turbulent confining
magnetic field leads to a destruction of the magnetic flux surfaces
\citep{Rosenbluth:1966,Filonenko:1967} that prevent radial mixing of
hot central plasma with cold exterior plasma.  It was recognized that
collisional diffusion is unable to account for all of the electron
heat transport measured in experiments, and that magnetic field line
wander could potentially explain the additional, ``anomalous''
transport.  Further work quantitatively estimated the diffusion in
collisional and collisionless regimes, suggesting that fluctuations of
sufficient amplitude, caused by microinstabilities at the scale of the
ion gyroradius, would consistently explain both the stochastic nature
of the magnetic field and the observed electron heat
transport \citep{Rechester:1978}.

Exploring the role of magnetic field line wander in enhancing electron
heat transport in a tokamak plasma has continued over the years using
analytical calculations and test particle modeling, and with the
comparison of these results to experimental measurements
\citep{Galeev:1981,Krommes:1983,Haas:1986,Laval:1993,Spatschek:2008}.
Recent advancements in the direct numerical simulation of weakly
collisional plasma turbulence using nonlinear gyrokinetic
simulations \citep{Pueschel:2008,Nevins:2011,Wang:2011,Hatch:2012,Hatch:2013}
and other direct numerical approaches
\citep{del-Castillo-Negrete:2014,del-Castillo-Negrete:2016} have
enabled breakthrough studies of the cause of magnetic field line
wander and its impact on confinement in fusion plasmas. Under
fusion-relevant plasma conditions, gyrokinetic simulations showed that
the magnetic field indeed rapidly becomes stochastic through
gradient-instability-driven turbulence, but that this stochasticity
does not always produce a significant enhancement in the electron heat
flux \citep{Nevins:2011}. The development of stochasticity appears to
arise through nonlinear interactions among overlapping magnetic
islands \citep{Wang:2011}, supporting an idea proposed in early
studies based on analytical considerations \citep{Rechester:1978}. By
focusing on the properties of the dominant turbulent modes in
ion-temperature-gradient and trapped-electron-mode instability driven
turbulence, this overlapping of magnetic islands arises through the
nonlinear transfer of energy from the unstable ballooning modes of odd
parity to stable tearing modes of even parity
\citep{Hatch:2012,Hatch:2013}.

\subsection{The Cascade of Energy in Plasma Turbulence}
\begin{figure}
  \centerline{\hbox{ \hfill\resizebox{1.2\textwidth}{!}{\includegraphics{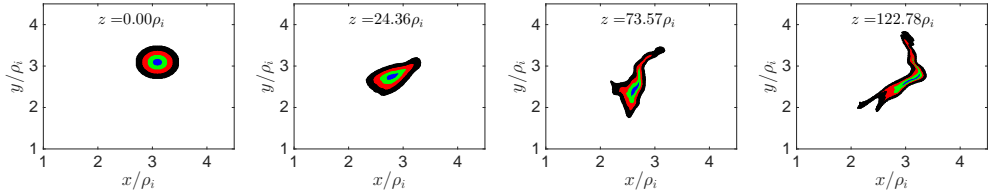}}\hfill }}
  \caption{ The distortion of a circular wavepacket as it propagates along the
    wandering magnetic field in a nonlinear gyrokinetic simulation of plasma turbulence.
    \label{fig:fluxtube} }
\end{figure}

It has been suggested that one can view the the cascade of energy to
small scales in magnetized plasma turbulence as due to the distortion
of the perpendicular structure of \Alfven wavepackets as they
propagating along a wandering, turbulent magnetic field
\citep{Similon:1989}, and this concept has been demonstrated
numerically \citep{Maron:2001}. An illustration of the distortion of a
wavepacket with an initially circular cross section perpendicular to
the equilibrium magnetic field from a nonlinear gyrokinetic simulation
of plasma turbulence is presented in figure~\ref{fig:fluxtube}.  Recent
work has adopted this framework to interpret current sheet formation
in coronal loops \citep{Rappazzo:2013}, the nonlinear transfer of
energy to smaller scale \Alfven waves in basic laboratory experiments
of plasma turbulence
\citep{Howes:2012b,Howes:2013a,Nielson:2013a,Howes:2013b,Drake:2013},
and the evolution of magnetic flux surfaces in 3D reduced MHD
simulations \citep{Servidio:2014b}.

\begin{figure}
\centerline{\resizebox{3.8in}{!}{\includegraphics{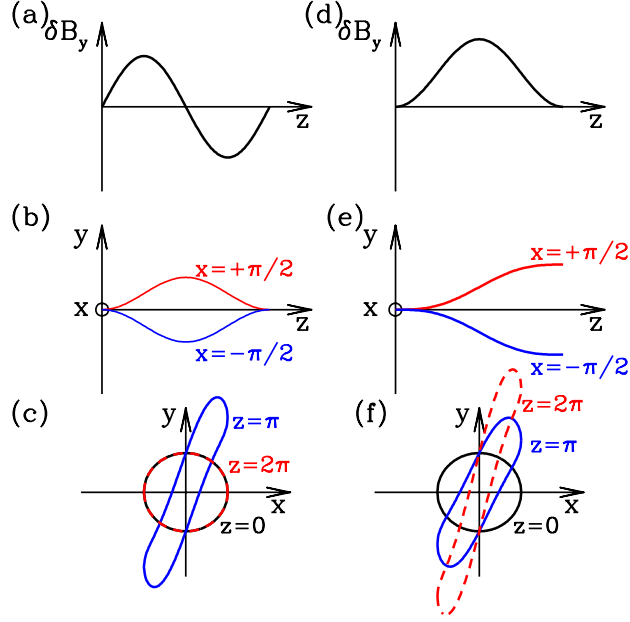}}}
\caption{Distortion of a magnetic flux tube in the
perpendicular $(x,y)$ plane due to wandering of the magnetic
field. (a) Magnetic field fluctuation $\delta B_y$ contains no $k_z=0$
component. (b) 3D tracing of magnetic field lines due to the magnetic
fluctuation in (a) at positions $x=+\pi/2$ (red) and $x=-\pi/2$
(blue). The oscillating shear causes the magnetic field to become sheared and then
subsequently to ``unshear.'' (c) The distortion of an \Alfven wave with an initially
circular perpendicular structure (black) as it propagates along the wandering
magnetic field. (d) For the case of a magnetic field fluctuation
$\delta B_y$ with a nonzero $k_z=0$ component, the (e) 3D magnetic field has a
monotonic shear, leading to (f) the permanent distortion (red) of an
initially circular \Alfven wave structure (black).
\label{fig:mag_shear}}
\end{figure}

Looking at this problem in more detail, the nonlinear physics
underlying the turbulent cascade of energy from large to small scales
is often described in Fourier space, where a nonlinear three-wave
coupling mechanism has been identified that leads to a secular
transfer of energy to modes with higher perpendicular wavenumber
\citep{Shebalin:1983,Sridhar:1994,Montgomery:1995,
  Ng:1996,Galtier:2000,Howes:2013a}, resulting in an anisotropic
cascade of energy in wavevector space.  But the physical manifestation
of the $k_\parallel =0$ mode that mediates this resonant three-wave
interaction is obscured by the use of the Fourier (plane-wave)
decomposition.  An analytical calculation modeling interactions
between localized \Alfven wavepackets demonstrated that wavepackets
involving a $k_\parallel=0$ component will lead to this lowest-order
three-wave nonlinear coupling \citep{Ng:1996}, and this mechanism has
been demonstrated using laboratory experiments
\citep{Howes:2012b,Howes:2013b,Drake:2013}.  Physically, the
$k_\parallel =0$ component of a wavepacket represents a shear in the
magnetic field, as depicted in
figure~\ref{fig:mag_shear}. Investigation of the propagation of
\Alfven waves in a wandering magnetic field demonstrates that the
cascade of energy to small scales is represented in physical space (as
opposed to Fourier space) as a shearing of the perpendicular structure
of an \Alfven wave as it propagates along a wandering magnetic field,
as depicted in figure~\ref{fig:mag_shear}(c) and (f).  Thus, exploring
the complementary picture of the plasma turbulent cascade as the
distortion of \Alfven waves as they propagate along a wandering
magnetic field may yield fresh insights into the nature of plasma
turbulence.

\subsection{Particle Acceleration}

Models of diffusive shock acceleration at collisionless shocks require
irregularities in the upstream magnetic field to return reflected
particles back to the shock front repeatedly to achieve significant
acceleration \citep{Ragot:2001,Guo:2010,Guo:2013,Guo:2015}.  The
efficiency of such a shock-acceleration mechanism is likely to be
dependent on the nature of the upstream magnetic field irregularities,
but existing investigations of these particle acceleration mechanisms
often use unrealistic models of the turbulent plasma, such as slab
turbulence with only a one-dimensional variation of the turbulent field,
$\delta \V{B}(z)$ \citep{Guo:2010}. The development of a new empirical
model of magnetic field line wander based on direct numerical
simulations of plasma turbulence will enable more realistic modeling
of the diffusive acceleration mechanism at collisionless shocks.

\subsection{Magnetic Reconnection}

It has been proposed theoretically that magnetic field line wander can
alter the physics of magnetic reconnection when the upstream plasma is
turbulent \citep{Lazarian:1999}, increasing the rate of reconnection
and the thickness of the current sheets in the turbulent medium
\citep{Vishniac:2012}. Numerical simulations have played a key role in
bearing out these ideas \citep{Kowal:2009,Kowal:2012}.  It has also
been claimed that exponential field line separation in turbulent
plasmas will lead to reconnection even in the absence of intense
current sheets \citep{Boozer:2014}. It has been shown that
stochasticity of the magnetic field in MHD turbulence simulations
degrades the usual notion of flux-freezing in an MHD plasma,
potentially explaining fast reconnection of large-scale structures at
MHD scales \citep{Eyink:2013}. More recently, numerical simulations
have been used to explore how breaking field line connectivity by
stochasticity of the magnetic field can be a mechanism for fast
reconnection \citep{Huang:2014}.  The properties of the turbulent
upstream magnetic fields very likely influence how the reconnection
mechanism is modified, so an improved model of magnetic field line
wander in plasma turbulence will contribute to progress in the
understanding of magnetic reconnection under the turbulent conditions
of realistic space and astrophysical plasma environments.

\section{Theoretical Considerations}
\label{sec:theory}

This study aims to use direct numerical simulations to illuminate
the physical processes influencing the development and saturation of
magnetic field line wander in astrophysical plasma turbulence.  The
ultimate goal is to construct an empirical description of the magnetic
field line wander in terms of the fundamental turbulence and plasma
parameters. This empirical description can be utilized to describe
accurately the properties of the turbulently tangled magnetic field
for application to studies of energetic particle transport in
astrophysical and laboratory plasmas, the cascade of energy to small
scales in plasma turbulence, the acceleration of particles to high
energies by collisionless shocks, and the physics of magnetic
reconnection in a turbulent plasma.

\subsection{Improving Our Understanding through Direct Numerical Simulations}

As detailed above, the nature of the magnetic field line wander in
previous studies appears to be quite sensitive to the characteristics
and parameters of the models describing the turbulent magnetic
field. Many of the models of the turbulent magnetic field used in the
literature exploring the phenomenon of magnetic field line wander and
its implications are severely outdated, being inconsistent with the
current understanding of plasma turbulence.  Specifically, models
employing slab, 2D, composite (slab plus 2D), and isotropic
distributions of magnetic fluctuations, although more susceptible to
theoretical analysis, are at odds with the anisotropic, 3D nature of
plasma turbulence that is now well-established through decades of
experimental, analytical, and numerical research on magnetized plasma
turbulence
\citep{Robinson:1971,Belcher:1971,Zweben:1979,Montgomery:1981,
  Shebalin:1983,Cho:2000,Maron:2001,Cho:2004,Cho:2009,TenBarge:2012a}. Recent
direct multi-spacecraft measurements of turbulence in the solar wind
confirm this anisotropic nature of the turbulent fluctuations
\citep{Sahraoui:2010b, Narita:2011,Roberts:2013,Roberts:2015b}.

In addition, the models employed in the studies reviewed above almost
universally employ randomly phased magnetic field fluctuations, a
characteristic inconsistent with any self-consistent realization of a
turbulent magnetic field.  Kolmogorov's Four-fifths
Law \citep{Kolmogorov:1941b} is an exact statistical formula relating
the mean energy dissipation rate to the third-moment of the velocity
fluctuations in hydrodynamic turbulence; subsequently, this
third-moment approach has been extended for the case of MHD turbulence
\citep{Chandrasekhar:1951,Politano:1998a,Yousef:2007}. A spectrum of
magnetic fluctuations with random phases yields an average
third-moment of zero, because it is the phase correlations among
different magnetic fluctuations that are responsible for the nonlinear
turbulent cascade of energy. Random-phase models therefore lack some
of the inherent attributes of a self-consistently determined turbulent
magnetic field \citep{Howes:2015b,Howes:2016b}.  It remains an open question whether such correlations
will indeed alter the nature of the magnetic field line wander
resulting from plasma turbulence, but the apparent sensitivity of the
results of previous studies to the characteristics of the magnetic
field model suggests that a self-consistent turbulent magnetic field
will yield the most well-justified and physically relevant results.

The uncertainty in  describing the turbulent magnetic field can be
eliminated by using a turbulent magnetic field that is generated
self-consistently by direct numerical simulation of the equations
governing the turbulent plasma dynamics. Of the studies directly
investigating magnetic field line wander reviewed above, only the
recent studies of stochastic magnetic field development in fusion
plasmas \citep{Pueschel:2008,Nevins:2011,Wang:2011,Hatch:2012,Hatch:2013} employed
direct numerical simulations, specifically using a nonlinear gyrokinetic
code.  A general approach to understand the development and
saturation of the tangled magnetic field in general astrophysical
plasma turbulence has not been attempted, and this provides a strong
motivation for the present work.

\subsection{The Role of Magnetic Reconnection}

\begin{figure}
  \centerline{\hbox{ \hfill\resizebox{3.2in}{!}{\includegraphics{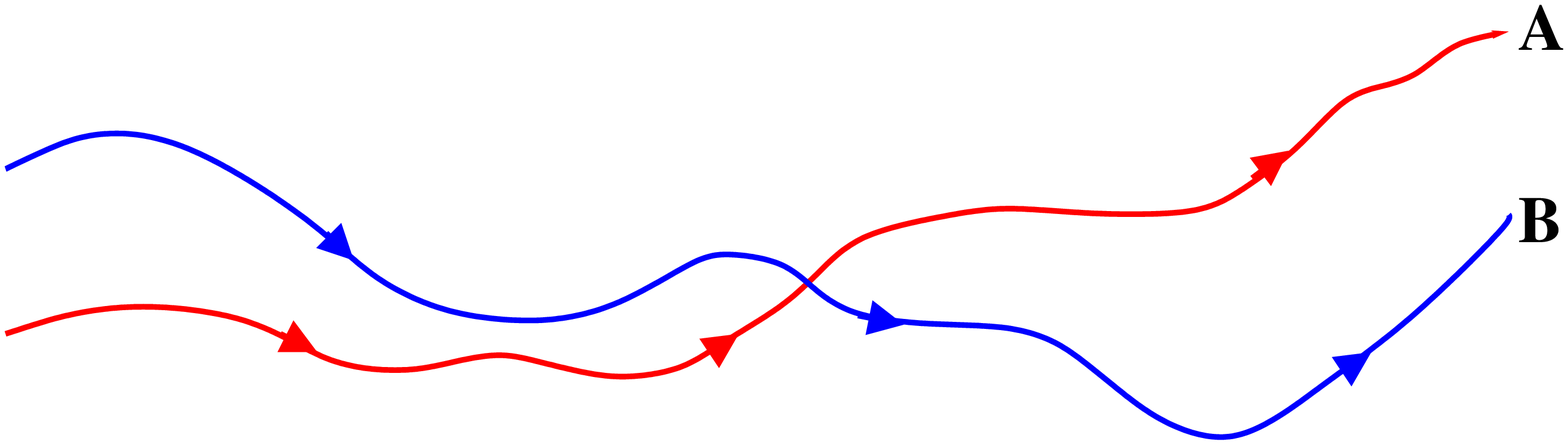}}\hfill }}
    
  \centerline{\hbox{  \hfill\resizebox{3.2in}{!}{\includegraphics{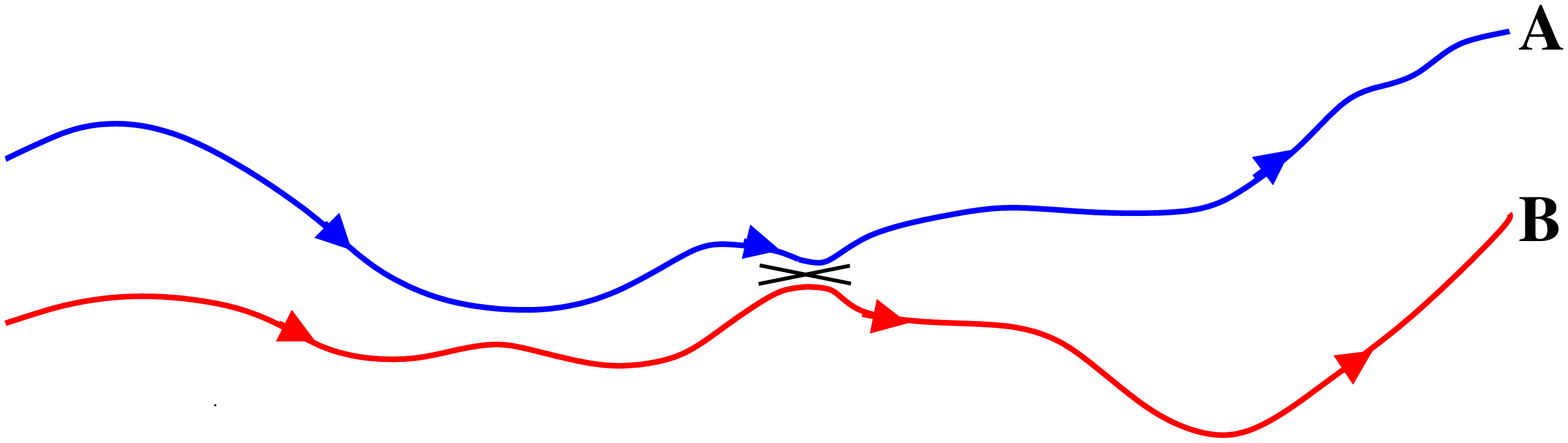}} \hfill }}
  \caption{ Illustration of how magnetic reconnection can
    instantaneously change the magnetic connectivity in a turbulent
    plasma, thereby influencing the development of magnetic field line
    wander.
    \label{fig:rxn} }
\end{figure}

Another important ingredient in understanding how magnetic fields
become dynamically tangled by plasma turbulence is the process of
magnetic reconnection. Although the impact of pre-existing turbulent
magnetic fields on the process of magnetic reconnection has been
examined previously
\citep{Lazarian:1999,Kowal:2009,Kowal:2012,Vishniac:2012,Boozer:2014,Huang:2014},
the investigation of how reconnection mediates the tangling and
untangling of magnetic fields in plasma turbulence has not been
addressed by any existing study. For example, consider the two
turbulent magnetic field lines depicted in figure~\ref{fig:rxn}.  When
a small reconnection event (denoted by the black cross) occurs between
the two field lines, the connectivity of those two field lines changes
instantaneously. Particles streaming along the red field line from the
left end of the plasma, which previously had been connected to point
A, will suddenly find themselves magnetically connected instead to
point B. An open question is whether the impact of magnetic
reconnection on the wandering of magnetic field lines can be described
in the framework of a diffusion, or whether this possibility of sudden
jumps in connectivity alter the nature of the resulting magnetic field
line wander. Direct numerical simulations of the turbulent plasma that
resolve the physics of magnetic reconnection, even in the
collisionless limit relevant to many space and astrophysical systems
of interest, are critical for answering this open question.

\subsection{The Turbulent Solar Wind as a Fiducial Example}

The turbulent solar wind that pervades our heliosphere represents a
fiducial system supporting a broad spectrum of turbulent plasma
motions over more than seven orders of magnitude in scale
\citep{Kiyani:2015}. It is also the most thoroughly diagnosed
turbulent astrophysical plasma in the universe, thereby providing
unique opportunities for confronting any empirical description of
magnetic field line wander with direct, \emph{in situ} measurements of
the turbulent interplanetary magnetic field. Thus, the solar wind
provides an ideal case study for discussing the phenomenon of magnetic
field line wander in a turbulent plasma. In addition, the tangling of
the magnetic field in the solar wind has important consequences for
the propagation of solar energetic particles erupting from violent
solar activity, constituting an important space weather hazard for
spaceborne human and technological assets.  Improving the modeling of
the tangled magnetic field in the heliosphere therefore has important
societal implications.

\begin{figure}
\centerline{\resizebox{\textwidth}{!}{\includegraphics*[0.35in,5.3in][8.1in,9.6in]{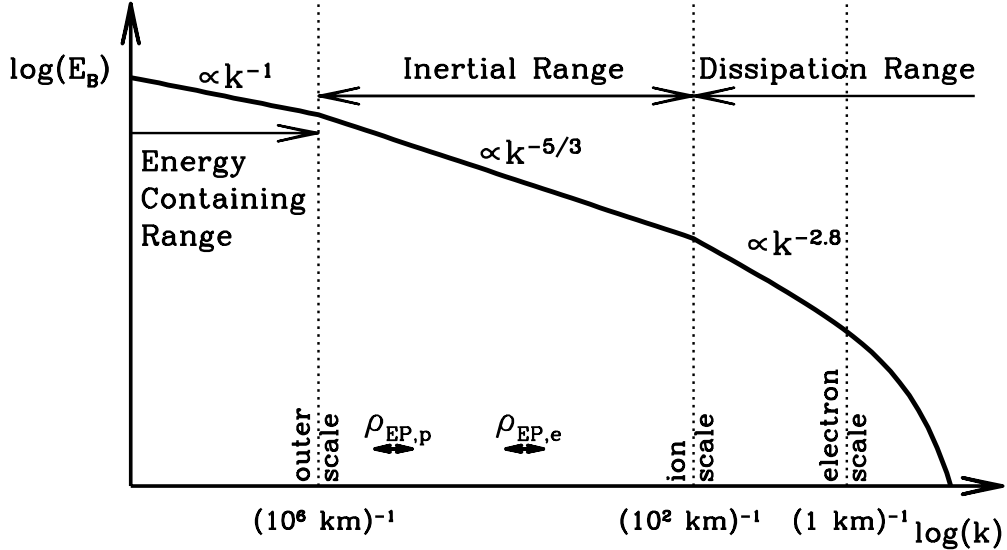}}}
\caption{Schematic of the magnetic energy wavenumber spectrum in the
  solar wind, showing the form of the spectrum in the energy
  containing, inertial, and dissipation ranges. Ranges for the typical
  Larmor radius scales for protons $\rho_{EP,p}$ and electrons
  $\rho_{EP,e}$ from Solar Energetic Particle (SEP) events are
  depicted.
\label{fig:spectrum}}
\end{figure}

Figure~\ref{fig:spectrum} presents a diagram of the solar wind
magnetic energy wavenumber spectrum---assuming the Taylor hypothesis
\citep{Taylor:1938} to convert the spacecraft-frame frequency to a
corresponding wavenumber of spatial fluctuations in the
super-\Alfvenic solar wind \citep{Howes:2014a,Klein:2014b}---for
near-Earth space at $R \sim 1$~AU.  At the largest scales $l >
10^6$~km (lowest wavenumbers) is the \emph{energy containing range}
\citep{Matthaeus:1994a,Tu:1995,Bruno:2005}, populated by large-scale
plasma flow and magnetic field fluctuations.  Through nonlinear
interactions, these energy containing fluctuations feed their energy
into the turbulent cascade at the outer scale, $l \sim 10^6$~km, where
the steepening of the magnetic energy spectrum marks the beginning of
the turbulent \emph{inertial range} \citep{Tu:1995,Bruno:2005}. Within
the inertial range, energy is nonlinearly transferred scale by scale
in a self-similar manner, leading to a single power-law spectrum down
to the inner scale, which corresponds to one of the characteristic ion
kinetic length scales at $l\sim 10^2$~km.  At this ion scale, the
magnetic energy spectrum breaks once again \citep{Bourouaine:2012}, marking the transition to
the \emph{dissipation range}\footnote{The name dissipation range is
  the most commonly used term, although use of this term is \emph{not}
  intended to imply that the steepening of the spectrum here is
  necessarily due to dissipation.}
\citep{Sahraoui:2009,Kiyani:2009,Alexandrova:2009,Chen:2010b,Sahraoui:2010b,Sahraoui:2013b}. Finally,
at the characteristic electron length scale of $l\sim 1$~km, the
magnetic energy spectrum is often observed to exhibit an exponential
roll-off \citep{Alexandrova:2012,Sahraoui:2013b}, interpreted to
indicate the ultimate termination of the turbulent
cascade. Also plotted on figure~\ref{fig:spectrum} is a representation of the
Larmor radius scales for protons and electrons from Solar Energetic
Particle (SEP) events, $\rho_{EP,p}$ and $\rho_{EP,e}$; collisionless
wave-particle interactions of energetic particles with the turbulent
magnetic fluctuations lead to scattering rates that peak at these
Larmor scales. Simulations using the gyrokinetic code \T{AstroGK} reproduce
quantitatively the features of the solar wind magnetic energy spectrum
from the middle of the inertial range down to the sub-electron scales
\citep{TenBarge:2012b,TenBarge:2013b}.

It is the turbulent magnetic field fluctuations over this broad range
of scales that lead to the stochastic character of the magnetic field,
so investigating how fluctuations in different scale ranges of this
spectrum affect the magnetic field line wandering is an important
long-term goal.  To accomplish this goal, direct numerical simulations
can be used to learn how magnetic field line wander develops and
saturates in plasma turbulence.

\subsection{A Model for the Development and Saturation of Magnetic Field Line Wander}

Here we propose a novel theoretical picture of the plasma physical
processes involved in the development of magnetic field line wander
and its saturation. The nonlinear interactions that underlie the
turbulent cascade of energy to small scales also lead to a tangling of
the magnetic field.  In an ideal plasma with no magnetic reconnection,
this tangling of the field would continue indefinitely, generating
turbulent structures on ever smaller scales, leading to an ever more
intricate wandering of the magnetic field.  But, once the turbulent
structures have reached sufficiently small scales that non-ideal
physics breaks the frozen-in condition, magnetic reconnection may
ensue, thereby untangling the magnetic field. We propose that the
saturation of the magnetic field line wander represents a balance between
the nonlinearly-driven tangling and the magnetic-reconnection-mediated
untangling, a physical picture that we aim to test thoroughly using
direct numerical simulations.

A key observation is that these two mechanisms, nonlinear interactions
(turbulence) and magnetic reconnection, depend differently on the
fundamental parameters describing the turbulence and the plasma.  For
example, the physics of magnetic reconnection has a strong dependence
on the plasma $\beta$, but the physics of the turbulent nonlinear
energy transfer appears to have very little dependence on this plasma
parameter \citep{Howes:2013a,Nielson:2013a,Howes:2015b}. What is
entirely new here, compared to the body of literature on turbulent
magnetic field line wander, is that most previous works\footnote{With
  the exception of studies directly focusing on magnetic
  reconnection.}  have completely ignored any role played by magnetic
reconnection.  From the theoretical picture above---a balance between
turbulent tangling and reconnective untangling---we believe that it is
inevitable that magnetic reconnection plays an important role in the
physics of magnetic field line wander, possibly leading to an
important, and as yet unrecognized, dependence on the plasma
$\beta$.\ A study based on direct numerical simulations that resolve
the kinetic microphysics of magnetic reconnection is likely to break
new ground on this important frontier in the study of magnetic
field line wander and its implications for many important physical
processes in the universe.

Discussions of the tangling of the magnetic field by plasma turbulence
often employs the term ``stochastic magnetic field'' as a generic
label for any turbulently tangled magnetic field. Here we reserve the
use of the term \emph{stochastic} to cases in which the magnetic field
indeed demonstrates a stochastic character, as demonstrated by an
appropriate analysis, such as a Poincare recurrence plot (see
section~\ref{sec:poincare}).  We choose to use the term \emph{magnetic
  field line wander} to refer to the general case when the magnetic
field does exhibit a topology in which the separation between two
field lines increases (or decreases) as one moves a distance along the
field, whether or not the magnetic field exhibits a stochastic
character.  How the physics of energetic particle propagation in
astrophysical plasmas, heat and particle transport in fusion plasmas,
magnetic reconnection, and particle acceleration differs when the
magnetic field lines merely wander, but are not fully stochastic, is
interesting open question.

\subsection{Key Questions}
To investigate and characterize the development and saturation of
magnetic field line wander in turbulent plasmas, direct numerical
simulations provide a valuable tool. Since magnetic reconnection may
play an essential role in the physics of the field line tangling, a
numerical method that resolves the kinetic microphysics of magnetic
reconnection, particularly under the weakly collisional conditions
relevant to many turbulent space and astrophysical plasma systems, is
essential. This can be supplemented by fluid methods to explore how
the nonlinear dynamics at large scales governs the tangling of the
magnetic field. Such an approach, maintaining a strong connection to
the nonlinear dynamics of plasma turbulence and magnetic reconnection,
will enable important new questions about the nature of magnetic field
line wander caused by plasma turbulence to be addressed:
\begin{enumerate}
\item How does magnetic field line wander develop and saturate in plasma turbulence?
\item Do the properties of magnetic field line wander depend on the underlying physical mechanism that enables magnetic reconnection (collisionless vs. resistive vs. numerical reconnection)?

\item Is magnetic field line wander dominated by the large-scale or small-scale fluctuations of the turbulence?

\item Can we construct an empirical description of the magnetic field line wander in terms of the  fundamental  plasma and turbulence parameters?
\end{enumerate}

\subsection{Quantitative Dependence on Fundamental Turbulence and Plasma Parameters}

Turbulence in heliospheric and other astrophysical plasmas naturally
leads to the tangling of the magnetic field, leading to the
development of magnetic filed line wander, a  property that we would
like to characterize quantitatively. To develop a quantitative measure
of the wandering of field lines in such a plasma, consider choosing
two points on different field lines separated by a distance $\delta r$
perpendicular to the magnetic field.  For these two particular points,
one may compute the perpendicular separation between the two field
lines in the perpendicular plane as a function of the distance along
the magnetic field line $l$. Computing statistics of this quantity
using direct numerical simulations of plasma turbulence enables the
development of the quantitative characterization of the magnetic filed
line wander.

The statistics of this perpendicular separation $\delta r$ between two
magnetic field lines can be quantitatively characterized in terms of a
small number of important parameters.  First are basic parameters to
describe the wandering of field lines away from each other, including
the distance traveled along the field line $l$ and the time for which
turbulence has been dynamically tangling the field.  Next are the
dimensionless parameters that describe the turbulence itself.  The
amplitude, or strength, of the turbulence is characterized by the
\emph{nonlinearity parameter}, $\chi=k_\perp \delta
B_\perp/(k_\parallel B_0)$ \citep{Goldreich:1995,Howes:2013a},
describing the amplitude of the turbulence at the driving scale
(equivalent to the ``Kubo'' number in some previous studies
\citep{Zimbardo:2000}). For example, to explore the impact on particle
diffusion by the amplitude of turbulent fluctuations, one study
computed particle trajectories numerically in a tangled magnetic field
consisting of 1000 randomly phased plane-wave modes
\citep{Hauff:2010}, and another study used Ramaty High Energy Solar
Spectroscopic Imager (RHESSI) observations of hard X-rays to estimate
the magnitude of magnetic field-line diffusion in flaring coronal
loops under several assumptions \citep{Bian:2011}. Both studies
found that, in the large Kubo number limit ($\chi \gg 1$), perpendicular diffusion
becomes independent of the turbulence spectrum, in agreement with the
predictions of percolation theory
\citep{Gruzinov:1990,Isichenko:1992}.

Another key dimensionless parameter is the \emph{isotropic driving
  wavenumber}, $k_0\rho_i$ \citep{Howes:2008b,Howes:2011b}, describing
the driving scale, or energy injection scale, of the turbulence
normalized to the ion Larmor radius, where it is assumed the turbulent
fluctuations are isotropic with respect to the direction of the mean
magnetic field at this scale\footnote{The assumption of isotropic
  driving may be relaxed if physical arguments suggest anisotropic
  driving, where an appropriate scaling theory for the anisotropic
  cascade of plasma turbulence can be used to devise a suitable
  dimensionless parameter to replace $k_0\rho_i$.}. Finally, the key
dimensionless parameters that describe the plasma are the ion plasma
beta, $\beta_i = 8 \pi n_0 T_i/B_0^2$, and the ion-to-electron
temperature ratio, $T_i/T_e$.  In summary, therefore, we expect that
the separation between two magnetic field lines $\delta r$ as one
follows the field lines through the plasma to have a dependence on
these fundamental parameters given by $\delta r(l,t,\chi,k_0\rho_i,
\beta_i, T_i/T_e)$.

For the most likely case of saturated turbulence in steady state,
where the tangling of the magnetic field by the turbulence and
untangling of the field by magnetic reconnection statistically
balance, we conjecture that the $\delta r$ will not depend on time.  A
reasonable form to seek for the average squared field line separation
in a turbulent steady state, $\langle(\delta r)^2\rangle$, based on
the previous analyses reviewed in section~\ref{sec:impact}, can be
written
\begin{equation}
  \langle(\delta r)^2\rangle = A \ l^{p}.
  \label{eq:dr2}
\end{equation}
Here $A(\chi,k_0\rho_i, \beta_i, T_i/T_e)$ represents the amplitude of
the average spreading of field lines, and we assume the average
squared field line separation can be expressed as a power law of the
distance along the field line $l$, given by the exponent
$p(\chi,k_0\rho_i, \beta_i, T_i/T_e)$.  The long term goal of this
project is to determine empirically forms for $A$ and $p$ that can be
used to characterize the magnetic field line wander in terms of the
turbulence and plasma parameters.

Although the two turbulence parameters $\chi$ and $k_0\rho_i$ together
with the two plasma parameters $\beta_i$ and $T_i/T_e$ leads to a four
dimensional parameter space, the broad range of different turbulence
models previously used to explore magnetic field line wander requires
a much longer list of possible parameters. 
The most likely culprit responsible for the many conflicting results
found in the literature is the tremendous variation among the
different models chosen to describe the turbulent magnetic field. The
use of direct numerical simulations of plasma turbulence enables us to
eliminate the huge parameter space necessary to describe the plethora
of different magnetic field models reviewed in
section~\ref{sec:impact}, at the expense of the necessarily limited
dynamic range attainable by direct numerical simulations.  But we
believe that a more physically faithful characterization of the
magnetic field line wander in plasma turbulence can be patched
together through a judicious use of different simulation models
appropriate to different ranges of the turbulent spectrum, as depicted
in Figure~\ref{fig:spectrum}.
\section{Code and Description of Turbulence Simulations}
The Astrophysical Gyrokinetics Code, or \T{AstroGK}, described in
detail in \citet{Numata:2010}, evolves the perturbed gyroaveraged
distribution function $h_s(x,y,z,\lambda,\varepsilon)$ for each
species $s$, the scalar potential $\varphi$, parallel vector potential
$A_\parallel$, and the parallel magnetic field perturbation $\delta
B_\parallel$ according to the gyrokinetic equation and the
gyroaveraged Maxwell's equations \citep{Frieman:1982,Howes:2006}, where
$\parallel$ is along the total local magnetic field $\V{B}=B_0 \zhat+
\delta \V{B}$. The velocity space coordinates are
$\lambda=v_\perp^2/v^2$ and $\varepsilon=v^2/2$. The domain is a
periodic box of size $L_{\perp }^2 \times L_z$, elongated along the
equilibrium magnetic field, $\V{B}_0=B_0 \zhat$. Note that, in the
gyrokinetic formalism, all quantities may be rescaled to any parallel
dimension satisfying $\epsilon \equiv L_\perp /L_z \ll 1$. Uniform
Maxwellian equilibria for ions (protons) and electrons are used, with
the correct mass ratio $m_i/m_e=1836$. Spatial dimensions $(x,y)$
perpendicular to the equilibrium field are treated pseudospectrally;
an upwind finite-difference scheme is used in the parallel
direction. Collisions are incorporated using a fully conservative,
linearized Landau collision operator that includes energy diffusion
and pitch-angle scattering due to electron-electron, ion-ion, and
electron-ion collisions \citep{Abel:2008,Barnes:2009}, yielding an
isotropic Maxwellian stationary solution.

The simulations presented here are similar to previous nonlinear
gyrokinetic simulations used to investigate the physics of the
turbulence in the weakly collisional solar wind
\citep{Howes:2008a,Howes:2011a}, in particular the small scale
simulations of the kinetic \Alfven wave cascade down to the scales of
the electron Larmor radius \citep{TenBarge:2013a,TenBarge:2013b}.  We
present here results of a simulation of a driven, strongly turbulent
kinetic \Alfven wave cascade in a plasma with parameters $\beta_i =1$
and $T_i/T_e =1$, where $\beta_i =v_{ti}^2/v_A^2$, $v_A$ is the
Alfv\'{e}n speed, $v_{ti} = \sqrt{2T_i/m_i}$ is the ion thermal speed,
and temperature is expressed in units of energy. The simulation domain
has dimensions
$(n_x,n_y,n_z,n_\lambda,n_\varepsilon,n_s)=(64,64,32,32,16,2)$,
yielding a simulation covering the fully dealiased range of $5 \le
k_\perp \rho_i \le 105$, or $0.12 \le k_\perp \rho_e \le 2.5$. It is
worthwhile noting that it has been demonstrated, via comparisons with
PIC simulations, that nonlinear gyrokinetic simulations using
\T{AstroGK} accurately describe the physics of magnetic reconnection
in the strong guide field limit, as long as the simulation resolves
the electron Larmor radius scale \citep{TenBarge:2014b}.

The simulation is driven at the domain scale using an oscillating
Langevin antenna \citep{TenBarge:2014a} to achieve turbulence with a
nonlinearity parameter $\chi \sim 1$, yielding critically balanced,
strong turbulent cascade of kinetic \Alfven waves
\citep{Goldreich:1995,Howes:2008b,Howes:2011b,TenBarge:2014a}.  The
linear frequency of a kinetic \Alfven wave at the domain scale,
determined by the collisionless linear gyrokinetic dispersion relation
\citep{Howes:2006}, is given by $\omega = 3.6 \omega_{A0}$, where
$\omega_{A0} = k_{\parallel 0} v_A$.  The parameters of the
oscillating Langevin antenna are amplitude $A_0=0.2$, driving
frequency $\omega_0=3.6\omega_{A0}$, and decorrelation rate
$\gamma_0=0.6\omega_{A0}$. We drive four modes, $(k_{\perp 0} \rho_i,
k_{\perp 0} \rho_i, k_{\parallel 0}\rho_i /\epsilon) = (5,0,\pm 1)$
and $(0,5,\pm 1)$, where $k_{\perp 0} = 2 \pi /L_\perp$ and
$k_{\parallel 0} = 2 \pi /L_z$. Note that the 3D simulation spatial
domain has size $L_\perp^2 \times L_z$, with $L_\perp = 2 \pi\rho_i
/5$ and $L_z = 2 \pi \rho_i /(5 \epsilon)$, where $\epsilon \ll 1$ is
the arbitrary gyrokinetic expansion parameter.  These parameters are
found to yield a statistically steady-state value of the nonlinearity
parameter of $\chi \simeq 1$.

Collision frequencies $\nu_i = 0.2 \omega_{A0}$ and $\nu_e = 0.5
\omega_{A0}$ are chosen to prevent a build-up of small-scale structure
in velocity space, yet to avoid altering the weakly collisional
dynamics, $\nu_s \ll \omega$.  All dissipation required to achieve a
statistically steady state in the simulation occurs via physically
resolved interactions, primarily collisionless damping via the Landau
resonance with electrons; no additional hypercollisionality is needed
in this simulation.

\section{Development and Saturation of Turbulent Cascade and Magnetic Field Line Wander}
In this section, we present numerical results describing the
development and saturation of the turbulent magnetic energy spectrum
and of the magnetic field line wander.  In addition, we use Poincare
plots to characterize the development of stochasticity in the magnetic
field.
\subsection{Development and Saturation of Turbulent Magnetic Energy Spectrum}
\begin{figure}
\centerline{\includegraphics[width=0.6\textwidth]{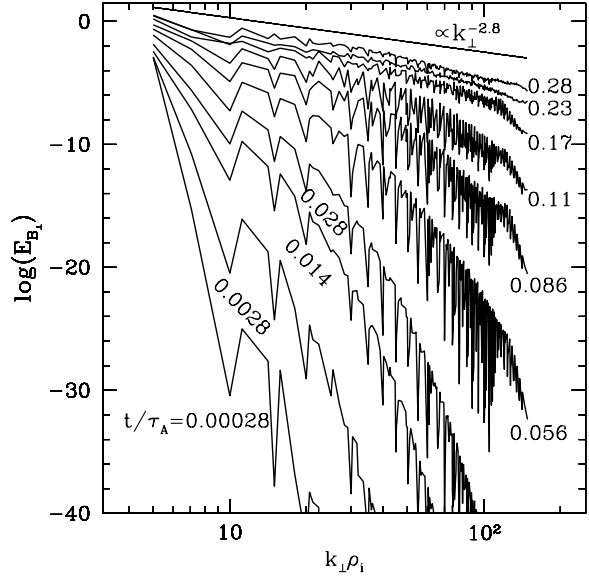}}
\caption{Development of the magnetic energy spectrum in the nonlinear
  gyrokinetic simulation of driven kinetic \Alfven wave turbulence.
  The normalized time for each spectrum $t/\tau_A$ is labeled.  The
  saturated spectrum for strong kinetic-\Alfven-wave cascade is
  expected to have a spectral index of $-2.8$, plotted for
  comparison.}
\label{fig:earlydevelop}
\end{figure}

These driven turbulence simulations begin with a straight, uniform
magnetic field and zero fluctuations.  The oscillating Langevin
antenna drives an external parallel current that generates
perpendicular magnetic field fluctuations, each driven mode with a
plane-wave pattern specified by the wavevector. Nonlinear interactions
between the counterpropagating \Alfven waves generated by the antenna
immediately begin to transfer energy to higher wavenumbers, and the
magnetic energy spectrum begins to fill in as energy is continually
injected into the driven modes. Figure~\ref{fig:earlydevelop} presents
the perpendicular magnetic energy spectrum $E_{B_\perp}(k_\perp) =
\int_{-\infty}^\infty d k_z \int_0^{2 \pi} d\theta \ k_\perp |\delta
B(k_\perp)|^2/8 \pi$ at a number of times early in the development of
the turbulent energy spectrum, where time is normalized using the
linear kinetic \Alfven wave period for modes at the domain scale,
$t/\tau_A$ and $\omega_{A0}\tau_A =1.74$. Note that, due to the broad
range of the logarithmic vertical axis, at the very early times
$t/\tau_A< 0.1$, very little energy has been injected into the
turbulent cascade.

\begin{figure}
\centerline{\includegraphics[width=0.6\textwidth]{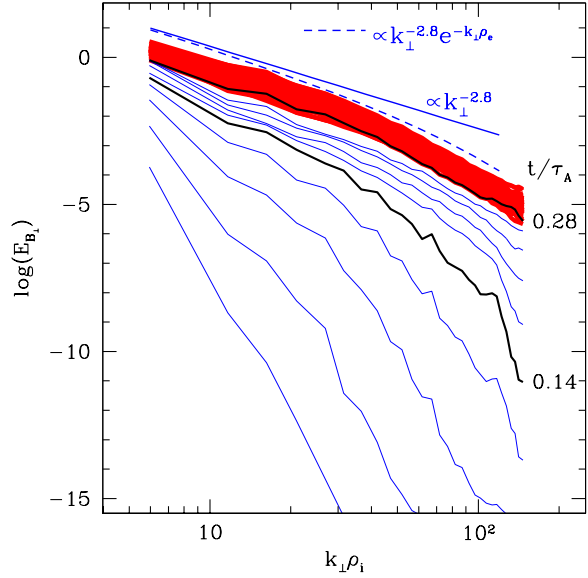}}
\caption{Perpendicular magnetic energy spectra plotted at uniform time
  intervals of $\delta t/\tau_A=0.028$ (blue and black), and all
  spectra overplotted over the range $0.28 \le t/\tau_A \le 3.6$
  (red).  For comparison, the slope for a power-law spectrum
  $E_{B_\perp} \propto k_\perp^{-2.8}$ (blue solid) and an
  exponentially cutoff spectrum $E_{B_\perp} \propto k_\perp^{-2.8}$
  (blue dashed) are shown for comparison.}
\label{fig:saturation}
\end{figure}

Figure~\ref{fig:saturation} shows how the spectrum saturates to a
statistically steady state.  Ten perpendicular magnetic energy spectra
(blue and black) are plotted at uniform linear time intervals
$t_j/\tau_A= j \Delta t/\tau_A$ for $j = 1, 2, 3, \ldots, 10$ and
$\Delta t/\tau_A=0.028$. We also overplot all spectra over the time
range $0.28 \le t/\tau_A \le 3.6$ (red). This plot shows that,
although, more energy is injected into the turbulent magnetic energy
spectrum at $t/\tau_A > 0.28$, the shape of the spectrum appears to be
saturated at $t/\tau_A = 0.28$---only the total energy content
changes, but the shape of the spectrum does not.  We also plot the
exponentially cutoff magnetic energy spectrum (blue dashed) determined
empirically from a sample of 100 \emph{Cluster} spectra,
$E_{B_\perp}(k_\perp) \propto (k_\perp \rho_i)^{-2.8} \exp( - k_\perp
\rho_e)$ \citep{Alexandrova:2012}, a result reproduced previously
using nonlinear gyrokinetic simulations
\citep{TenBarge:2013a,TenBarge:2013b}. The present simulation also
agrees well with this model magnetic energy spectrum.

Note that the spectra in Figure~\ref{fig:saturation} have been binned
in bins of width $\Delta k_\perp \rho_i =5$, producing a smoother
appearance than the spectra in Figure~\ref{fig:earlydevelop}, in which
all possible values of $ k_\perp \rho_i$ from our perpendicular
wavevector grid are represented.

The timescales associated with the development and saturation of the
turbulent cascade can be nicely illustrated by plotting the amplitude
of the energy at a perpendicular wavenumber in the middle of the
dynamic range at $k_\perp \rho_i=20.6$ as a function of time, as shown
in Figure~\ref{fig:twotimescales}(a).  At early times $t/\tau_A
<0.14$, the perpendicular magnetic energy at $k_\perp \rho_i=20.6$
increases as a steep power law with time, $E_{B_\perp}(k_\perp)
\propto t^{12}$. At time $t_1/\tau_A =0.14$, the increase of energy
with time at this wavenumber reduces to a less steep approximate power
law with $E_{B_\perp}(k_\perp) \propto t^{6}$, before reaching a
statistically steady value at $t_2/\tau_A = 0.28$.\footnote{Note that
  the power law scaling of the amplitude reported here is dependent on
  the value of $k_\perp \rho_i$ chosen.  A detailed study of how this
  scaling varies with $k_\perp \rho_i$ is beyond the scope of the work
  presented here.} Beyond this time, the shape of the magnetic energy
spectrum no longer evolves, as illustrated by the spectra plotted in
Figure~\ref{fig:saturation}, but the total amplitude of the energy
varies over a factor of four, with the same spectral shape simply
shifting up and down.  This variation of the total energy in the
spectrum is due to the fact that the finite-time-correlated driving of
the oscillating Langevin antenna leads to a significant variation in
the rate of energy injection \citep{TenBarge:2014a}.

It is worthwhile pointing out how rapidly the magnetic energy spectrum
saturates, requiring a length of time that is only a fraction of the
period of the kinetic \Alfven wave at the domain scale, $0.28 \tau_A$.
This is likely due to the dispersive nature of the kinetic \Alfven
wave fluctuations in the sub-ion length scale regime. $k_\perp \rho_i
\gg 1$, where the parallel phase velocity of the waves increases
linearly with $k_\perp \rho_i$, with an approximate scaling
\citep{Howes:2014a}
\begin{equation}
\frac{\omega}{k_\parallel}= v_A \frac{k_\perp\rho_i}{\sqrt{\beta_i + 2/(1+T_e/T_i)}}.
\label{eq:vphase}
\end{equation}
Thus, the increasingly fast dynamics of the fluctuations with
decreasing scale appears to saturate the energy spectrum more rapidly
than the wave period of the domain scale kinetic \Alfven waves that
drive the cascade.  It is also worthwhile noting that the turbulent
cascade is very efficiently generated by driving counterpropagating
kinetic \Alfven waves at the domain scale.

In Figure~\ref{fig:twotimescales}(b), we plot the evolution of the
spectral index $\eta$ of the perpendicular magnetic spectrum
$E_{B_\perp}$ as a function of normalized time $t/\tau_A$.  To
determine the value of the spectral index, we perform a fit of the 
perpendicular magnetic spectrum at each time using the form
\begin{equation}
E_{B_\perp} \propto (k_\perp \rho_i)^\eta \exp[-k_\perp \rho_e]
\end{equation}
over the range of scales $10 \le k_\perp \rho_i \le 105$.  As shown in
panel (b), the rate of increase of the spectral index decreases,
particularly for $t >t_1$, finally saturating to a constant
value for $t \ge t_2$.

\subsection{Development and Saturation of Magnetic Field Line Wander}
To estimate the development and saturation of magnetic field line
wander, we devise a new diagnostic, the \emph{expansion parameter}
$\sigma$, derived in Appendix~\ref{app:expansion}. This diagnostic
yields a scalar quantity that parameterizes the separation of magnetic
field lines in the turbulent magnetic field throughout the simulation domain.

\begin{figure}
\centerline{\includegraphics[width=0.6\textwidth]{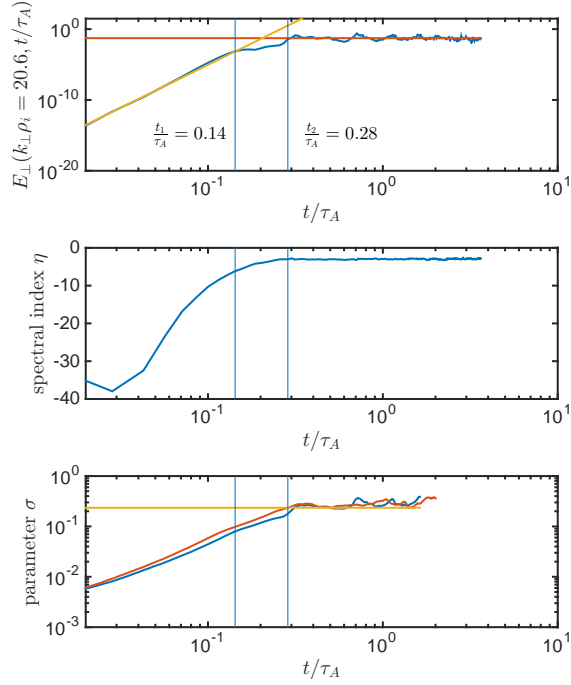}}
\caption{(a) Evolution of the energy in modes with $k_\perp
  \rho_i=20.6$ vs.~normalized time $t/\tau_A$, showing power-law
  growth until $t_1/\tau_A=0.14$, and a slower growth until
  $t_2/\tau_A=0.28$ (b) Evolution of the spectral index, $\eta$,
  where the spectrum is fit by $(k_\perp \rho_i)^\eta \exp(-k_\perp \rho_e)$,
  showing saturation at time $t_2/\tau_A=0.28$. (c) Evolution of the
  scalar expansion parameter $\sigma$, showing that the separation of
  field lines saturates only at $t_2/\tau_A=0.28$, with little
  noticeable change at $t_1/\tau_A=0.14$. A second identical run
  (with a different pseudo-random number sequence governing the
  forcing) shows the evolution is statistically repeatable.}
\label{fig:twotimescales}
\end{figure}

In Figure~\ref{fig:twotimescales}(c), we plot the evolution of
expansion parameter $\sigma$ as a function of normalized time
$t/\tau_A$. We plot the results for two independent simulations (red
and blue), with all of the same numerical and physical parameters, but
different pseudo-random number sequences governing the driving by the
oscillating Langevin antenna.  The expansion parameter $\sigma$
increases as a power law right up to the time $t_2/\tau_A=0.28$, at
which point $\sigma$ saturates to a statistically steady value. Note
that unlike in panel (a), where the rate of increase of amplitude of
turbulent cascade shows a marked decrease at time $t =t_1$, the
power-law increase of the expansion parameter $\sigma$ in panel (c)
shows little change at $t =t_1$. Note also that the two
independent simulations generate statistically similar results for the
evolution of this expansion parameter $\sigma$.  Therefore, it appears
for these simulations with plasma parameters $\beta_i =1$ and $T_i/T_e
=1$ and turbulent amplitude $\chi \simeq 1$, the magnetic field line
separation appears to saturate on the same timescale as the turbulent
magnetic energy spectrum. 

\section{Development of Magnetic Stochasticity}
\label{sec:poincare}
In well developed turbulence, the magnetic field appears to become
tangled up in a stochastic manner, raising two very important
questions about the development of magnetic stochasticity in turbulent
plasmas.  First, how long must the turbulence evolve before the
magnetic field becomes stochastic? Second, must the amplitude of the
turbulence exceed some threshold value for the development of
stochasticity?  We reiterate here that we reserve the term
``stochastic magnetic field'' for a field topology that displays a
stochastic nature in a Poincare plot (see below), whereas we employ
the term ``magnetic field line wander'' for the general case of a
turbulent magnetic field, whether or not that field demonstrates a
stochastic character.

Although Figure~\ref{fig:twotimescales}(c) shows the timescale of the
saturation of the separation of magnetic field lines, as diagnosed by
the scalar expansion parameter $\sigma$, it does not provide alone any
information about whether the magnetic field has become stochastic.
To investigate the development of stochasticity, we use Poincare plots
to yield a qualitative measure of the stochasticity
\citep{Dombre:1986,Wang:2011,Nevins:2011}.

To construct the Poincare plot, we begin with the magnetic field
$\V{B}(\V{x})$ at some time $t$. On the perpendicular plane at one end
of the simulation domain at $z=0$, we specify a sparse pattern of
points with the color of each point creating a bullseye pattern.  The
magnetic field line passing though each point is traced through the
domain to the far end of the simulation domain at $z=L_z$, and a point
is plotted there, with color matching that of the original field line
position.  That field line is periodically wrapped to $z=0$, and the
process is continued, with a colored point plotted at each crossing at
$z=L_z$.  We trace through the box $20$ times for each field line,
generating a sufficient number of points in the Poincare plot to
visually determine whether or not the magnetic field has become
stochastic.  A fourth-order Runge-Kutta method with adaptive step-size
is used to trace each field line by integrating the ordinary
differential equation $d \V{r}/dl = \bhat(\V{x})$, where $\bhat(\V{x})
= \V{B}(\V{x})/|\V{B}(\V{x})|$.  If the field line passes through the
boundaries of the simulation domain in the $x$ or $y$ directions, it
is periodically wrapped to the opposite boundary. We have checked that
using the field-line following routine to trace back down the field
line returns to the original starting point.

\begin{figure}
  \centerline{\resizebox{\textwidth}{!}{\includegraphics{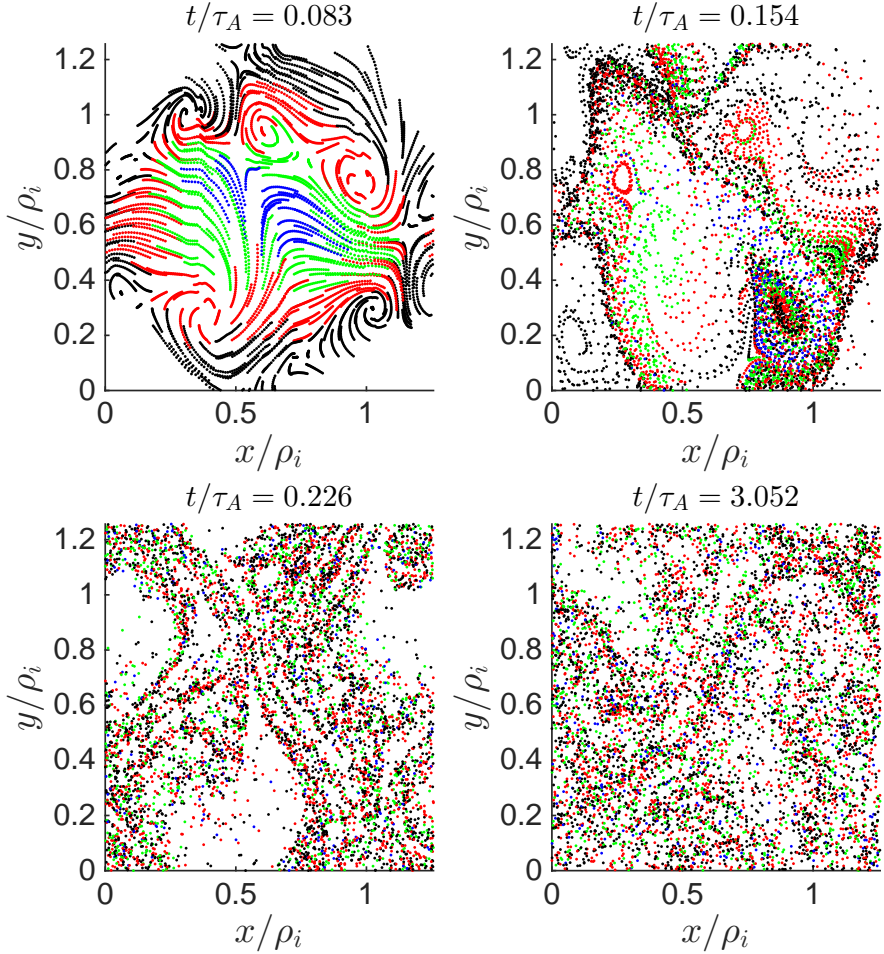}}}
  \caption{Poincare plots in the $z=L_z$ plane that diagnose the
    magnetic field topology at times $t/\tau_A=0.083$, $0.154$,
    $0.226$, and $3.052$, showing regions of stochasticity at
    $t/\tau_A=0.154$ and fully developed stochasticity for $t/\tau_A
    \gtrsim 0.226$.}
\label{fig:poincare}
\end{figure}

Figure~\ref{fig:poincare} shows the Poincare plots for the magnetic
field at times $t/\tau_A=0.083$, $0.154$, $0.226$, and $3.052$.  At
the early time $t_1/\tau_A=0.083$, the Poincare plot remains well
ordered, indicating that the magnetic field has not yet become
stochastic.  By the time $t/\tau_A=0.155$, some regions of the
Poincare plot demonstrate a disordered mixture of colored points,
indicating regions that have become stochastic, while other regions
maintain some semblance of order.  Thus, it appears that as the
magnetic field becomes stochastic, that stochasticity manifests itself
in some regions of the domain but not others. By time
$t/\tau_A=0.226$, the entire domain demonstrates a stochastic
character.  Since the turbulent spectrum appears to saturate to a
constant shape at $t_2/\tau_A=0.28$, we conclude that, for this
critically balanced simulation with nonlinearity parameter $\chi \sim
1$, the magnetic field is generically stochastic for well developed
turbulence.  Since space and astrophysical plasmas are almost always
found to be turbulent, this finding has significant implications for
many systems of interest.

Note that we can use these results to relate indirectly the expansion
parameter $\sigma$ to the development of stochasticity of the magnetic
field.  The expansion parameter has the advantage that it can be
computed locally, depending only on the value of the magnetic field
and its derivatives at a single point, whereas the computation of the
Poincare plot requires knowledge of the magnetic field throughout the
simulation domain. Future work will more thoroughly compare the
expansion parameter $\sigma$ to Poincare plots for cases of turbulence
with different amplitudes to determine whether the $\sigma$ can be
useful as a proxy to estimate the stochasticity of a given magnetic field
configuration.

\vspace{0.2cm} 
\section{Conclusion}
Here we have presented a general context for understanding the
tangling of magnetic lines in a turbulent plasma from direct numerical
simulations of weakly collisional plasma turbulence.  We have
discussed the broader issue of magnetic field line wander, including the
effect on the propagation of cosmic rays and energetic particles in
the heliosphere and other astrophysical plasmas, the degradation of
particle confinement in magnetic fusion devices, the relation to the
cascade of energy in plasma turbulence, and the role played by tangled
magnetic fields in particle acceleration and magnetic reconnection.
We have identified some key questions about the development and
saturation of magnetic field line wander in plasma turbulence, and
discussed the fundamental turbulence and plasma parameters upon which
this behavior is likely to depend.

We present the analysis of a driven kinetic \Alfven wave simulation of
strong plasma turbulence with plasma parameters $\beta_i =1$ and
$T_i/T_e =1$ and a nonlinearity parameter $\chi \simeq 1$.  We show how the
magnetic energy spectrum develops in time from uniform conditions and
saturates to a steady spectral shape, and identify two timescales that
characterize the evolution and saturation.  Next, we investigate the
development of tangling in the magnetic field using a scalar expansion
parameter $\sigma$ to characterize the separation of field lines in
the plasma turbulence. Finally, we use Poincare plots to qualitatively
demonstrate how the magnetic field attains a stochastic character in
our strong plasma turbulence simulation.

We find that, for the case of strong plasma turbulence analyzed here,
the magnetic field indeed develops a stochastic character.  We expect
that, on general grounds, the magnetic field in turbulent space and
astrophysical plasmas is likely to always be stochastic. Such a
stochastic nature of the magnetic field is very important to the
prediction of the propagation of energetic particles through the
turbulence heliospheric interplanetary magnetic field, with
implications for energetic particles hazards to spaceborne robotic and
human assets.

The companion work to this paper, \citet{Bourouaine:2016}, explores
the magnetic field line wander over a range of turbulent amplitudes
$0.1 \lesssim \chi \lesssim 5$ with $\beta_i=1$ and $T_i/T_e=1$ to
test whether one should indeed always expect turbulent astrophysical
plasmas to contain a stochastic magnetic field. We find that the
magnetic field becomes fully stochastic when the turbulence amplitude
exceeds a threshold value, $\chi > \chi_{\mbox{thresh}} \simeq
0.1$. Analysis of the spreading of the field lines finds slightly
superdiffusive behavior for stronger turbulence with $\chi > 1$ and
slightly subdiffusive behavior for weaker turbulence with $\chi < 1$,
and we provide a functional form for the dependence on the turbulent
amplitude $\chi$ for the coefficient $A(\chi)$ and exponent $p(\chi)$
in \eqref{eq:dr2}.  For the case of critically balanced turbulence
with $ \chi \sim 1$, the behavior appears to be quite close to
diffusive, with the exponent in \eqref{eq:dr2} given by $p(\chi=1)
\simeq 1$.  This appears to be in agreement with the
Rechester-Rosenbluth model of diffusive behavior
\citep{Rechester:1978}, although away from critical balance the direct
numerical simulations suggest this simple model may break down.

\appendix
\section{Definition of the Scalar Expansion Parameter $\sigma$}
\label{app:expansion}
It is useful to derive a scalar quantity that can parameterize the
separation of field lines in a turbulent magnetic field. For this
purpose, we derive here the scalar \emph{expansion parameter},
$\sigma$.

We begin with the magnetic field specified throughout the triply
periodic simulation domain at a single time, $\V{B}(\V{x})$. We define
the unit vector that specifies local direction of the magnetic field,
\begin{equation}
  \bhat(\V{x})\equiv \frac{\V{B}(\V{x})}{|\V{B}(\V{x})|},
  \label{eq:bhat}
\end{equation}
where $\V{x}=(x,y,z)$ is the position.  Field-line tracing can be
performed by following along the local magnetic field direction at
each point,
\begin{equation}
  \frac{\partial \V{r}}{\partial l}= \bhat(\V{r}),
\end{equation}
where $l$ is the distance along the magnetic field line.  A field line
is defined by the vector $\V{r}(l,\V{r}_0)$, where the starting point
at $l=0$ is $\V{r}_0=\V{r}(0,\V{r}_0)$.

Since we are interested in the separation between field lines as you
progress along either of those field lines, we chose another field
line $\V{r}'$ separated from the field line $\V{r}$ by the separation
$\delta \V{r}$, such that $\V{r}'(l,\V{r}'_0)=\V{r}(l,\V{r}_0)+ \delta
\V{r}(l)$. Taylor expanding the field line $\V{r}'$ about $\V{r}$, one
may obtain an expression for the evolution of $\delta \V{r}$ as you
move a distance $l$ along the field line $\V{r}$ ,
\begin{equation}
\frac{\partial \delta \V{r}}{\partial l}= (\delta \V{r} \cdot \nabla)
\bhat(\V{r})
  \label{eq:separation}
\end{equation}

Since we are primarily interested here in the separation of magnetic
field lines due to turbulence at scales sufficiently below the outer
scale of the inertial range, we take the local magnetic field to be
$\V{B}(\V{x}) = B_0 \zhat + \delta \V{B}(\V{x})$, where the
perturbations are small compared to the mean magnetic field, $|\delta
\V{B}| \ll B_0$. In this limit, the total magnetic field magnitude
\begin{equation}
  B= |\V{B}| = \sqrt{B_0^2 + 2 B_0 \delta B_z + |\delta \V{B}|^2} \simeq
  B_0 + \delta B_z  + \mathcal{O}(|\delta \V{B}|^2)
\end{equation}
Substituting this result into the definition of $\bhat$ and dropping
terms of order $\mathcal{O}(|\delta \V{B}|^2/B_0^2)$ and higher, we
obtain a first-order expression for the magnetic field direction,
\begin{equation}
\bhat(\V{r}) = \zhat + b_x \xhat + b_y \yhat,
\end{equation}
where $b_x = \delta B_x /B_0$ and $b_y = \delta B_y /B_0$.  Therefore,
to first order, the variation of the direction of the magnetic field
depends only on the components of the magnetic field perpendicular to
the mean magnetic field, $\V{B}_0=B_0 \zhat$. In this limit, to first
order in $|\delta \V{B}|$, we may express the displacement of the
magnetic field lines $\delta \V{r}(l)$ in terms of the initial
displacement $\delta \V{r}_0$ by
\begin{equation}
  \delta \V{r}(l) = \alpha(l) \mathsfbi{R}[\phi(l)] \cdot \delta \V{r}_0,
  \label{eq:sepmodel}
\end{equation}
where $\alpha(l)$ represents expansion (positive) or contraction
(negative) of the separation between the field lines, and
$\mathsfbi{R}[\phi(l)]$ is a matrix representing rotation \emph{in the
  perpendicular plane} of the vector by an angle $\phi(l)$,
\begin{equation}
  \mathsfbi{R}[\phi(l)] \equiv \left( \begin{array}{rr}
    \cos [\phi(l)] & - \sin [\phi(l)] \\
      \sin [\phi(l)] & \cos [\phi(l)] \\
\end{array}\right).
\end{equation}
Thus, the evolution of the displacement (in the perpendicular plane,
to lowest order) between two particular magnetic field lines may be
characterized by the two scalar quantities $\alpha(l)$ and $\phi(l)$.

Here we are primarily interested in the separation of field lines
$\alpha(l)$, so we eliminate $\phi(l)$ by taking a dot product of
\eqref{eq:sepmodel} with itself, obtaining the result,
\begin{equation}
  |\delta \V{r}|^2 = \alpha^2 | \delta \V{r}_0|^2
  \label{eq:alpha_relation}
\end{equation}

We can obtain an expression for $\alpha$ by using \eqref{eq:separation}
to obtain
\begin{equation}
 \delta \V{r}= \delta l (\delta \V{r}_0 \cdot \nabla)
\bhat(\V{r}_0) + \delta \V{r}_0,
\end{equation}
an expression for the separation valid to first order in $\delta l$.
One may then obtain the expression, valid to first order in $\delta
l$, 
\begin{equation}
\alpha =\frac{ |\delta \V{r}| }{ | \delta \V{r}_0|}= 1 + \frac{ \delta \V{r}_0  \cdot [(\delta \V{r}_0 \cdot \nabla)
    \bhat(\V{r}_0) ]}{ | \delta \V{r}_0|^2} \delta l.
\label{eq:alpha}
\end{equation}

Finally, we define the dimensionless scalar \emph{expansion parameter}, $\sigma$, by
\begin{equation}
  \frac{\partial (\delta r/\rho_i)}{\partial (l/a_0)}= \sigma (\delta r_0/\rho_i)
\end{equation}
where the perpendicular length scale is normalized by the ion thermal
Larmor radius $\rho_i = v_{ti}/\Omega_i = (2 T_i/m_i)^{1/2}m_i c/q_i
B_0$ and the parallel length scale is normalized by a characteristic
length $a_0$.  Note that by normalizing to two separate length scales,
this definition enables a strong connection to gyrokinetic simulations
in which the gyrokinetic expansion parameter, $\epsilon \equiv
\rho_i/a_0 \sim|\delta \V{B}|/B_0 \ll 1$, enables the results of a
single simulation to be scaled to any ratio for the strength of the
equilibrium field to the perturbed magnetic field, $B_0/ |\delta
\V{B}| \sim a_0/\rho_i$.  This expression can be simplified to
\begin{equation}
      \frac{\partial \delta r}{\partial l}=
      \sigma \left(\frac{\delta r_0}{\rho_i} \right)\left(\frac{\rho_i}{a_o}\right),
\end{equation}
where the presence of the gyrokinetic expansion parameter in the last
pair of parentheses yields a value of $\sigma$ of similar magnitude
for different ratios of $B_0/ |\delta \V{B}|$.

In the limit $\delta l \rightarrow 0$, this dictates $
\delta r = \delta r_0 + (\sigma \delta l/a_0) \delta r_0$. Comparing
this expression for $\alpha$ in \eqref{eq:alpha}, and using
\eqref{eq:alpha_relation}, we obtain a simple result for the expansion
parameter,
\begin{equation}
\sigma =  \delta \hat{\V{r}}_0  \cdot [(\delta \hat{\V{r}}_0 \cdot \hat{\nabla})
  \bhat(\V{r}_0) ],
\label{eq:sigmadef}
\end{equation}
where we define the direction of the initial separation vector $
\delta \hat{\V{r}}_0 = \delta \V{r}_0 /\delta r_0$. In addition, we
have normalized the gradient operator to the ion Larmor radius scale,
so that the perpendicular components of this operator (the only ones
that contribute to \eqref{eq:sigmadef}) are given by $
\hat{\nabla}_\perp = \xhat\rho_i( \partial/\partial x) + \yhat \rho_i
(\partial/\partial y)$.  If we take an initial separation
vector to have an angle $\gamma$ in the $x$-$y$ plane with respect to
the $x$-axis, $ \delta \hat{\V{r}}_0 = \cos \gamma \xhat + \sin \gamma
\yhat$, then this expression simplifies to
\begin{equation}
  \sigma =\cos^2 \gamma  \frac{\partial b_x }{\partial x}
  + \sin \gamma \cos \gamma \left(  \frac{\partial b_y }{\partial x} +  \frac{\partial b_x }{\partial y}\right) + \sin^2 \gamma \frac{\partial b_y}{\partial y}
\end{equation}

In the limit $|\delta \V{B}| \ll B_0$, one may obtain an expression
for $\bhat$ up to $\mathcal{O}(|\delta \V{B}|^2/B_0^2)$,
\begin{equation}
  \bhat = \frac{\V{B}}{|\V{B}|} =
  \zhat \left( 1 + \frac{|\delta \V{B}|^2}{B_0^2} \right)
  + \frac{\delta B_x}{B_0} \xhat + \frac{\delta B_y}{B_0} \yhat
  - \frac{\delta B_z}{B_0}  \frac{\delta \V{B}}{B_0} .
\end{equation}
Taking the divergence of the field direction, we obtain
\begin{equation}
  \nabla \cdot \bhat = 
   \frac{\partial}{\partial x} \left( \frac{\delta B_x}{B_0}\right)
  + \frac{\partial}{\partial y} \left( \frac{\delta B_y}{B_0}\right)
-\frac{\partial}{\partial x} \left(  \frac{\delta B_x \delta B_z}{B_0^2} \right)
-\frac{\partial}{\partial y} \left(  \frac{\delta B_y \delta B_z}{B_0^2} \right)
-\frac{\partial}{\partial z} \left(
\frac{\delta B_x^2\delta B_y^2}{B_0^2}
\right)
\end{equation}
In the anisotropic limit $\epsilon \equiv k_\parallel / k_\perp \sim
|\delta \V{B}| / B_0 \ll 1$ that is relevant to turbulent fluctuations
at small scales, then $\partial/\partial z \sim \epsilon
\partial/\partial x \sim \epsilon
\partial/\partial y$, and the terms in the equations above have the
following order: the first two terms  are $\mathcal{O}(\epsilon)$, the
next two terms are $\mathcal{O}(\epsilon^2)$, and the final term is
$\mathcal{O}(\epsilon^3)$.  Therefore, dropping terms of order
$\mathcal{O}(\epsilon^2)$ and higher, we obtain the simplified
expression,
\begin{equation}
  \nabla \cdot \bhat_\perp = 0,
\end{equation}
where $\bhat_\perp = \delta B_x/B_0 \xhat + \delta B_y/B_0 \yhat$.
Therefore, we obtain the important simplifying result, $\partial
  b_x /\partial x = - \partial b_y/\partial y$. This
important result means that, to lowest order, compressions and
expansions of the magnetic field are divergence free in the
perpendicular plane.

One consequence of this property is that, for a given initial field
line, $\V{r}$, the expansion parameter $\sigma$ yields a zero value
when integrated over all possible directions $\gamma$ of displacement about
that point. Specifically,
\begin{eqnarray}
 \int_0^{2\pi} d\gamma \sigma(\gamma)&  = &
   \int_0^{2\pi} d\gamma \left[ \cos^2 \gamma  \frac{\partial b_x }{\partial x}
     + \sin \gamma \cos \gamma \left(  \frac{\partial b_y }{\partial x} +  \frac{\partial b_x }{\partial y}\right) + \sin^2 \gamma \frac{\partial b_y}{\partial y}\right] \\
&   = &\frac{1}{2} \left( \frac{\partial b_x }{\partial x} +
    \frac{\partial b_y}{\partial y}\right)=0
\end{eqnarray}
Note that the values of the derivatives $\partial b_i /\partial x_j$
are constant in this integration.

The consequence of this finding is that the value of the expansion
parameter as a function of angle $\gamma$ is bounded by some maximum
value, $|\sigma(\gamma) | \le \sigma_{max}$.  Thus, at each point
$\V{r}$, we can simply use this maximum value as a scalar value that
simply characterizes the expansion (and the consequent equal and
opposite contraction at other angles necessary to yield $
\int_0^{2\pi} d\gamma \sigma(\gamma)=0$).  In the body of the paper,
the parameter $\sigma$ at a given instant of time is computed as the
value of $ \sigma_{max}$ averaged over all points in the simulation
domain. 

A treatment of the twisting of the magnetic field about itself,
characterized by the parameter $\phi(l)$, will be presented in
subsequent work.

The work has been supported by NSF CAREER Award AGS-1054061 and NASA
NNX10AC91G.

\bibliographystyle{jpp}


\end{document}